\documentclass[iop,apjl]{emulateapj}

\usepackage{graphicx}

\begin{document}
\title{A \emph{Spitzer}-IRS Detection of Crystalline Silicates in a Protostellar Envelope}

\author{Charles A. Poteet\altaffilmark{1}, S. Thomas Megeath\altaffilmark{1}, 
Dan M. Watson\altaffilmark{2},  Nuria Calvet\altaffilmark{3}, Ian S. Remming\altaffilmark{2},\\ 
Melissa K. McClure\altaffilmark{2,3}, Benjamin A. Sargent\altaffilmark{4}, William J. Fischer\altaffilmark{1},  Elise Furlan\altaffilmark{5}, Lori E. Allen\altaffilmark{6},\\ Jon E. Bjorkman\altaffilmark{1}, Lee Hartmann\altaffilmark{3}, James Muzerolle\altaffilmark{4},  John J. Tobin\altaffilmark{3}, and Babar Ali\altaffilmark{7}}

\altaffiltext{1}{Department of Physics and Astronomy, 
The University of Toledo, 2801 West Bancroft Street, Toledo, OH 43606; \email{charles.poteet@gmail.com}}

\altaffiltext{2}{Department of Physics and Astronomy, University of Rochester, 
Rochester, NY 14627}

\altaffiltext{3}{Department of Astronomy, University of Michigan, 830 Dennison Building, 500 Church Street, Ann Arbor, MI 48109}

\altaffiltext{4}{Space Telescope Science Institute, 3700 San Martin Drive, Baltimore, MD 21218}

\altaffiltext{5}{Jet Propulsion Laboratory, California Institute of Technology, Mail Stop 264Ð723, 4800 Oak Grove Drive, Pasadena, CA 91109}

\altaffiltext{6}{National Optical Astronomy Observatory, 950 North Cherry Avenue, Tucson, AZ 85719}

\altaffiltext{7}{NHSC/IPAC, California Institute of Technology, 770 South Wilson Avenue, Pasadena, CA 91125}

\slugcomment{Accepted to ApJ Letters: 2011 April 19}

\shorttitle{CRYSTALLINE SILICATES IN A PROTOSTELLAR ENVELOPE}
\shortauthors{POTEET ET AL.}

\begin{abstract}
 
We present the \emph{Spitzer Space Telescope} Infrared Spectrograph spectrum of the Orion~A protostar \object{HOPS-68}.  The mid-infrared spectrum reveals crystalline substructure at 11.1, 16.1, 18.8, 23.6, 27.9, and 33.6 $\micron$ superimposed on the broad 9.7 and 18 $\micron$ amorphous silicate features;  the substructure is well matched by the presence of the olivine end-member forsterite (Mg$_{2}$SiO$_{4}$).  Crystalline silicates are often observed as infrared emission features around the circumstellar disks of \object{Herbig Ae/Be} stars and \object{T Tauri} stars.  However, this is the first unambiguous detection of crystalline silicate absorption in a cold, infalling, protostellar envelope.  We estimate the crystalline mass fraction along the line-of-sight by first assuming that the crystalline silicates are located in a cold absorbing screen and secondly by utilizing radiative transfer models.  The resulting crystalline mass fractions of 0.14 and 0.17, respectively, are significantly greater than the upper limit found in the interstellar medium ($\lesssim$~0.02-0.05).  We propose that the amorphous silicates were annealed within the hot inner disk and/or envelope regions and subsequently transported outward into the envelope by entrainment in a protostellar outflow.    

\end{abstract}


\section{Introduction \label{sec:intro}}

Since their discovery in the envelopes around oxygen-rich evolved stars, crystalline silicates (e.g., pyroxene [(Mg, Fe)SiO$_{3}$] and olivine [(Mg, Fe)$_{2}$SiO$_{4}$]) have been detected in the circumstellar environments of both pre-- and post--main-sequence stars, comets, and ultraluminous infrared galaxies \citep[and references therein]{Henning10}. ÊHowever, silicates in the interstellar medium are almost entirely amorphous in structure \citep{Kemper05}.  Crystalline silicates are thought to be produced by thermal annealing or evaporation and re-condensation in warm environments; experimental studies indicate that such processes require temperatures of $\sim$ 800$-$1200 K \citep{Harker02,Gail04}. Ê

In pre--main-sequence stars, crystalline silicates are often detected in the warm inner disk region ($\lesssim$ 1~AU), where the temperature is sufficient to anneal amorphous silicates \citep[and references therein]{Bouwman08, Sargent09, Olofsson10}.  In contrast, there are only a few claims of crystalline silicates in protostars, which are still surrounded by cold, infalling envelopes and thus in an earlier evolutionary stage.  \citet{Ciardi05} attributed short-wavelength emission features in the protostellar binary \object{SVS 20} to crystalline silicates. ÊHowever, such emission must originate in the warm inner regions near the central protostar and may be produced in a circumstellar disk as in pre--main-sequence stars. ÊConversely, crystalline silicates in the cold, infalling envelope would result in absorption features.  Shallow absorption near 11.3 $\micron$ has been detected toward several low-mass protostars \citep{Boogert04,Riaz09}.Ê The association of this feature with crystalline silicates is debated: \citet{Li07} suggest instead that the inclusion of a H$_{2}$O ice mantle on silicate grains results in a weak shoulder near 11 $\micron$. ÊIn support of this, \citet{Riaz09} find a strong correlation between the 11.3 $\micron$ feature and the H$_{2}$O ice column density.

In this Letter, we present the \emph{Spitzer Space Telescope} \citep{Werner04} Infrared Spectrograph \citep[IRS;][]{Houck04} spectrum of the deeply embedded protostar \object{HOPS-68} \citep[also known as \object{FIR-2};][]{Mezger90}.  The source is located in the Orion Molecular Cloud 2 region \citep[OMC-2;][and references therein]{Gatley74,Peterson08}, northward of the Orion Nebula, at an adopted distance of 414 $\pm$ 7 pc \citep{Menten07}.  We report the first detection of a complex of \emph{absorption} features which can unambiguously be attributed to crystalline silicates in the cold envelope of a protostar.


\begin{figure*}[htp]
\centering
\rotatebox{90}{\includegraphics*[width=0.65\textwidth]{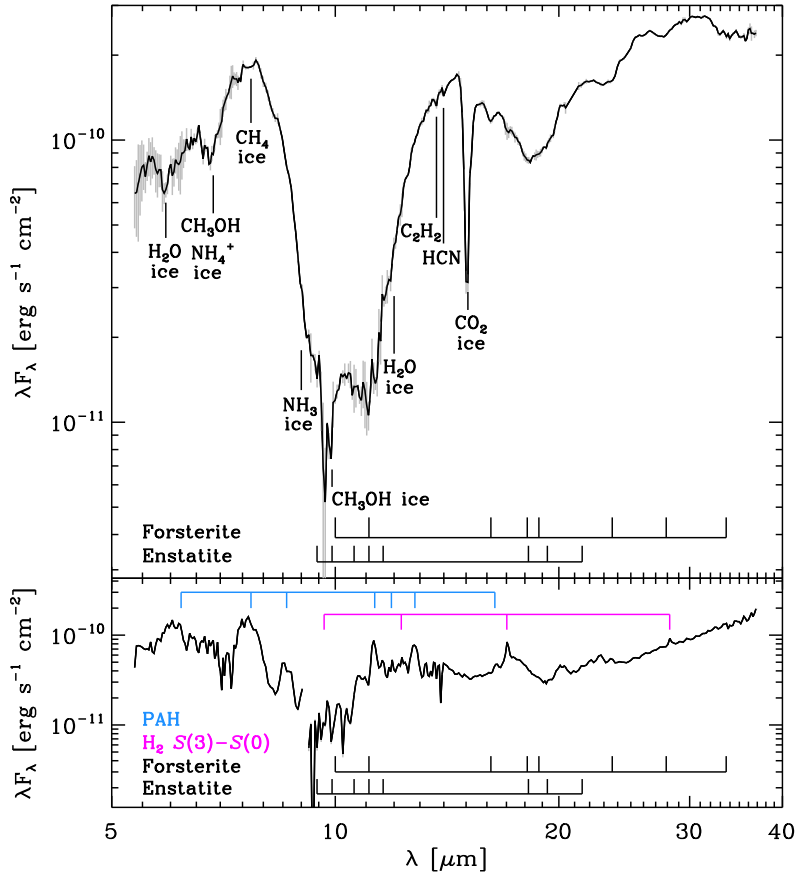}}
\caption{\emph{Top panel}: \emph{Spitzer}-IRS 5$-$36 $\micron$ spectrum of the protostar HOPS-68, with uncertainties (\emph{gray lines}) and spectral features indicated.  The feature at 9.66 $\micron$ is a background-subtraction artifact, arising from the H$_{2}$ \emph{S}(3) emission line near the source.  \emph{Bottom panel}:  The spectrum of the subtracted background.  Although PAH features and H$_{2}$ lines are detected, crystalline silicate features are not apparent.}
\label{fig:irs}
\end{figure*}


\section{Observations and Data Reduction \label{sec:obs}}

The 5-36 $\micron$ spectrum was obtained toward the \emph{Spitzer} 4.5 $\micron$ position of \object{HOPS-68} ($\alpha$~=~05$^{\rm{h}}35^{\rm{m}}24\fs3$, $\delta$~=~$-$05$\degr08\arcmin30\arcsec$ [J2000]) on 2007 March 27 (\emph{Spitzer} AOR 20838656) with the short- (SL; 5.2$-$14.0 $\micron$) and long-wavelength (LL; 14.0$-$36.1 $\micron$) low-resolution ($\lambda$/$\Delta\lambda$ = 60$-$120) IRS modules at each of the two nominal nod positions, one-third of the way from the slit ends.  The spectrum was extracted from the \emph{Spitzer} Science Center (SSC) S18.7 pipeline basic calibrated data using the Advanced Optimal Extraction in the SMART software package \citep{Higdon04} following the methods described in \citet{Lebouteiller10}.  The LL spectra and first order SL spectra were background-subtracted by fitting a degree zero polynomial to either side of the emission profiles.  For the SL module, the background emission was scaled up (down) by 30\% before taking the nod difference for the first (second) nod positions.

The resulting nods were then averaged to obtain the final spectrum, and the spectral uncertainties are estimated to be half the difference between the two independent spectra from each nod position.  We estimate the spectrophotometric accuracy of the final result to be 5\%.

Additional 70 and 160 $\micron$ imaging was performed on 2010 September 28 (observation IDs 1342205228 and 1342205229) with \emph{Herschel's} Photodetector Array Camera and Spectrometer \citep{Pilbratt10,Poglitsch10} as part of the \emph{Herschel} Orion Protostar Survey (HOPS).  An $8\arcmin$ field was repeatedly observed eight times using two orthogonal scanning directions and a scan speed of $20\arcsec$~s$^{-1}$, giving a total observation time of 3290~s.  The data were processed with ``Method 1'' described in \citet{Fischer10}.

Aperture photometry was obtained using a 9.6$\arcsec$ and 12.8$\arcsec$ aperture at 70 and 160 $\micron$, respectively, with the background annulus extending from 9.6$-$19.2$\arcsec$ and 12.8$-$25.6$\arcsec$, respectively.  An aperture correction was determined from the encircled energy fraction provided by the PACS consortium (private communication).  The photometric accuracy is dominated by calibration uncertainties, which are 10\% and 20\% at 70 and 160 $\micron$, respectively.

\section{Analysis \label{sec:ana}}

The \emph{Spitzer}-IRS 5$-$36 $\micron$ spectrum exhibits copious silicate and ice absorption features (Figure \ref{fig:irs}) that are typically found in the circumstellar environments of embedded protostars \citep{Boogert04, Watson04}.  The main ice constituents are H$_{2}$O (6.0 and 12.0 $\micron$), CH$_{3}$OH (6.8 and 9.9 $\micron$), CH$_{4}$ (7.7 $\micron$), NH$_{3}$ (9.0 $\micron$), and CO$_{2}$ (15.1 $\micron$).  In addition, the spectrum reveals two gas-phase organic molecules:  C$_{2}$H$_{2}$ (13.7 $\micron$) and HCN (14.0 $\micron$); these have been previously observed toward massive protostars \citep{An09}.  The prominent silicate absorption features at 9.7 and 18 $\micron$ are due to the  Si-O stretching and O-Si-O bending vibrational modes, respectively.  


\begin{figure}[htp]
\centering
\rotatebox{90}{\includegraphics*[width=1.08\textwidth]{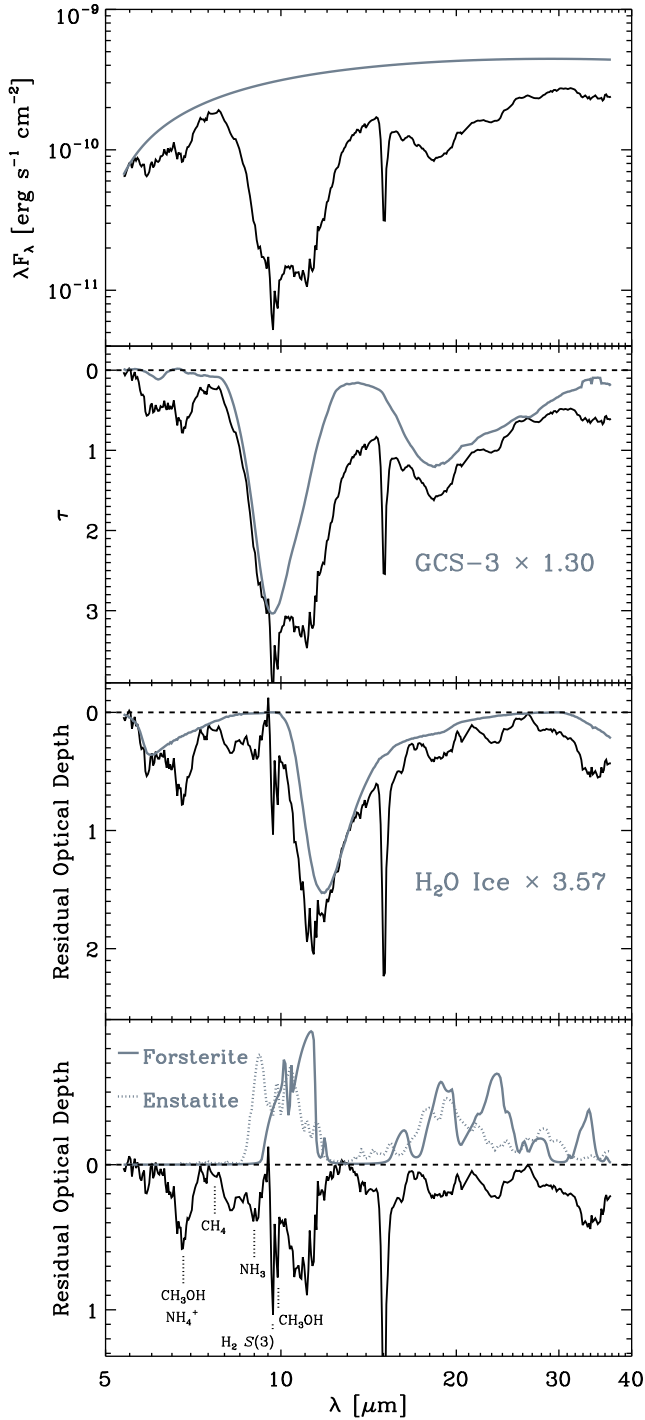}}
\caption{The isolation of crystalline silicate features in the spectrum of HOPS-68.  \emph{Top panel}:  The \emph{Spitzer}-IRS spectrum with the adopted second-order polynomial unabsorbed continuum (\emph{gray line}).  \emph{Second panel}:  The derived optical depth spectrum compared to the amorphous silicate profile of GCS-3 \citep[\emph{gray line};][]{Chiar06}.  \emph{Third panel}:  The residual optical depth spectrum, after subtracting the scaled amorphous silicate profile of GCS-3, compared to a laboratory spectrum of pure crystalline H$_{2}$O ice at \emph{T } = 140 K \citep[\emph{gray line};][]{Hudgins93}.  \emph{Bottom panel}:  The residual optical depth spectrum after subtracting the scaled H$_{2}$O ice profile.  Arbitrarily scaled mass absorption coefficients for forsterite (\emph{solid gray line}) and enstatite (\emph{dotted line}) are shown for comparison; these were calculated using optical constants from \citet{Sogawa06} and \citet{Jager98}, respectively.}
\label{fig:tau}
\end{figure}


The spectrum also reveals a complex of narrow features near 11, 16, 19, 24, 28, and 34 $\micron$.  These features are typically detected in emission around the disks of T Tauri stars \citep{Sargent09, Olofsson10} and have been attributed to the presence of crystalline silicates.  Moreover, this complex of features has previously only been detected in \emph{absorption} toward heavily obscured ultraluminous infrared galaxies \citep{Spoon06}.    

\subsection{Cold Absorbing Screen Method \label{sec:tau}}

To investigate the presence of crystalline silicate absorption, we approximate the envelope as a cold absorbing screen against a bright infrared continuum \citep[e.g.,][]{Boogert08}.  The continuum is approximated as a second-order polynomial that is fit to the regions of the spectrum that appear devoid of absorption features:  5.5, 7.5, and 31.6 $\micron$.  
 
The optical depth spectrum, $\tau(\lambda)$, is derived assuming $F_{\lambda}^{\rm{obs}}$ = $F_{\lambda}^{\rm{cont}}$$e^{-\tau(\lambda)}$, where $F_{\lambda}^{\rm{obs}}$ and $F_{\lambda}^{\rm{cont}}$ are the observed and continuum flux densities, respectively.  To account for the amorphous silicate contribution, the optical depth spectrum is compared to the silicate absorption profile of the Galactic Center source \object{GCS-3} \citep{Chiar06}, which is commonly used to represent ``standard'' interstellar silicate absorption.  The polynomial continuum fit was offset to a greater than observed flux at 7.5 and 31.6 $\micron$; the applied offsets ensured an observed optical depth greater than or equal to the combined \object{GCS-3} and H$_{2}$O ice optical depth for all wavelengths.  

The top panel of Figure \ref{fig:tau} shows the result after applying the offsets to the polynomial continuum fit.  The \object{GCS-3} optical depth was then constrained to fit the observed optical depth spectrum at 9.3$-$9.5 $\micron$ by minimizing 

\begin{equation}
\chi^{2} = \sum_{i} \left[\frac{\tau_{\rm{o}}(\lambda_{i}) - \alpha \tau_{\rm{m}}(\lambda_{i})}{\sigma_{\rm{o}}(\lambda_{i})}\right]^2,
\end{equation}

\noindent where $\tau_{\rm{o}}(\lambda_{i})$ and $\sigma_{\rm{o}}(\lambda_{i})$ are the observed optical depth and its corresponding uncertainty at the \emph{i}th wavelength, respectively, $\tau_{\rm{m}}(\lambda_{i})$ is the tabulated optical depth at the \emph{i}th wavelength for \object{GCS-3}, and $\alpha$ is the scaling parameter of the fit.  The scaled amorphous component is shown in the second panel of Figure \ref{fig:tau}.  Next, the residual optical depth spectrum is compared to the laboratory spectrum of pure crystalline H$_{2}$O ice at \emph{T} = 140 K \citep{Hudgins93}, which was calculated for spherical grains in the Rayleigh-limit (A.~C.~A. Boogert, private communication).  The H$_{2}$O ice profile is fit to match the residual optical depth spectrum between 12.6$-$13.0 $\micron$ (third panel of Figure \ref{fig:tau}).  The crystalline H$_{2}$O ice optical depth spectrum provides a good fit to the depth of the broad H$_{2}$O libration mode, and thus is preferred over the laboratory spectrum of pure amorphous H$_{2}$O ice.


\begin{deluxetable*}{ccccc}
\centering
\tablewidth{0pt}
\tablecolumns{5}

\tablecaption{Crystalline Mass Fraction \label{tbl:massfrac}}

\tablehead{
\colhead{$\lambda$} &
\colhead{$\kappa_{\rm{cr}}$($\lambda_{\rm{p}}$)\tablenotemark{a}} &
\colhead{$\tau_{\rm{cr}}$($\lambda_{\rm{p}}$)\tablenotemark{b}} &
\colhead{$N_{\rm{cr}}$/($N_{\rm{cr}}$+$N_{\rm{am}}$)\tablenotemark{c}} &
\colhead{$N_{\rm{cr}}$/($N_{\rm{cr}}$+$N_{\rm{am}}$)\tablenotemark{c}} \\
\colhead{[$\micron$]} & 
\colhead{[10$^{3}$ cm$^{2}$ g$^{-1}$]} &
\colhead{} &
\colhead{[Hudgins]} &
\colhead{[Warren]} 
}
\startdata
11.1 & 6.0 & 0.90 \tiny\phantom{$<$}(0.12) &  0.15 \tiny\phantom{$<$}(0.02) &  0.13 \tiny\phantom{$<$}(0.02) \\
16.1 & 1.5 & 0.26 \tiny\phantom{$<$}(0.01) &  0.16 \tiny\phantom{$<$}(0.01) &  0.20 \tiny\phantom{$<$}(0.01) \\
18.8 & 3.8 & 0.22 \tiny\phantom{$<$}(0.02) &  0.06 \tiny\phantom{$<$}(0.01) &  0.09 \tiny\phantom{$<$}(0.01) \\
23.6 & 4.2 & 0.23 \tiny($<$0.01) &  0.06 \tiny($<$0.01) & 0.07 \tiny($<$0.01)  \\
27.9 & 1.2 & 0.15 \tiny($<$0.01) &  0.13 \tiny($<$0.01) & 0.11 \tiny($<$0.01) \\
33.6 & 2.4 & 0.43 \tiny\phantom{$<$}(0.02) &  0.17 \tiny\phantom{$<$}(0.01) & 0.15 \tiny\phantom{$<$}(0.01) 
\enddata
\tablenotetext{a}{Peak mass absorption coefficients for forsterite.} 
\tablenotetext{b}{Peak optical depth of forsterite.}
\tablenotetext{c}{Crystalline mass fractions derived from Equation 2 using H$_{2}$O optical constants from \citet{Hudgins93} and \cite{Warren84}, respectively.}

\tablecomments{Uncertainties based on statistical errors in the \emph{Spitzer}-IRS spectrum only.  Peak optical depth of the 9.7 $\micron$ amorphous silicate feature measured from the GCS-3 fit:  $\tau_{\rm{am}}$($\lambda_{\rm{p}}$) = 3.05.}
\end{deluxetable*}


Subtraction of the scaled H$_{2}$O ice profile from the optical depth spectrum isolates features near 11, 16, 19, 24, 28, and 34 $\micron$, as illustrated in the bottom panel of Figure \ref{fig:tau}.  The peak positions, widths, and strengths of the residual features are inconsistent with the strongest features of the pyroxene end-member enstatite \citep[MgSiO$_{3}$;][]{Jager98}.  Conversely, the peak positions and widths of the residual features at 11.1, 16.1, 18.8, 23.6, 27.9, and 33.6 $\micron$ coincide with the strongest of those in the mass absorption coefficients (MACs) for forsterite \citep[Mg$_{2}$SiO$_{4}$;][]{Sogawa06}.  The relative strengths of the 18.8 and 23.6 $\micron$ features appear weaker than those calculated for forsterite; a similar mismatch between the observed and calculated strengths was reported by \citet{Spoon06}.  Nonetheless, we conclude that the features observed at 11.1, 16.1, 18.8, 23.6, 27.9, and 33.6 $\micron$ are caused by the presence of the mineral forsterite.   

The crystalline mass fractions are inferred from the peak optical depths of the 9.7 $\micron$ amorphous, $\tau_{\rm{am}}$($\lambda_{\rm{p}}$), and the 11.1, 16.1, 18.8, 23.6, 27.9, and 33.6 $\micron$ crystalline, $\tau_{\rm{cr}}$($\lambda_{\rm{p}}$), silicate absorption features using 

\begin{equation}
\frac{N_{\rm{cr}}}{N_{\rm{cr}}+N_{\rm{am}}} = \frac{\tau_{\rm{cr}}(\lambda_{\rm{p}})\kappa_{\rm{am}}(\lambda_{\rm{p}})}
{\tau_{\rm{cr}}(\lambda_{\rm{p}})\kappa_{\rm{am}}(\lambda_{\rm{p}})+\tau_{\rm{am}}(\lambda_{\rm{p}})\kappa_{\rm{cr}}(\lambda_{\rm{p}})},
\end{equation}

\noindent where $N_{\rm{cr}}$ and $N_{\rm{am}}$ are the dust mass column densities for crystalline and amorphous silicates, respectively.  The peak MACs for forsterite, $\kappa_{\rm{cr}}$($\lambda_{\rm{p}}$), are tabulated in Table \ref{tbl:massfrac} and were calculated using optical constants from \citet{Sogawa06} (\S\ref{sec:rt}).  Adopting $\kappa_{\rm{am}}$($\lambda_{\rm{p}}$) = 3.5 $\times$ 10$^{3}$ cm$^{2}$ g$^{-1}$ as the peak MAC for amorphous pyroxene \citep{Sargent06}, we find crystalline mass fractions ranging from 0.06$-$0.17, with a median value of 0.14.  These crystalline mass fractions are significantly greater than the upper limit found in the interstellar medium \citep[$\lesssim$ 0.02$-$0.05;][]{Kemper05,Li07}.  A factor-of-three variation in the crystalline mass fraction may be due to uncertainties in the MACs for forsterite, uncertainties in the subtracted optical depth spectra for the amorphous silicate and H$_{2}$O ice contributions, and finally the assumption that all the silicate material is in the cold absorbing screen.  Furthermore, assuming more realistic grain shapes for forsterite may reduce the discrepancies among the crystalline mass fractions, but we assume a continuous distribution of ellipsoids \citep[CDE;][]{Bohren83} for simplicity.  

The analysis is repeated using the H$_{2}$O ice optical constants from \citet{Warren84}.  The results show that the mass fraction of forsterite does not depend on the adopted H$_{2}$O ice optical constants (Table \ref{tbl:massfrac}).

\subsection{Radiative Transfer Method \label{sec:rt}}

To account for the presence of both warm emitting material in the inner envelope and cold absorbing material in the outer envelope, we now analyze the spectrum using a radiative transfer model.  The model is based on the method introduced in \citet{Kenyon93} and \citet{Calvet94}, and later refined in \citet{Osorio03} and \citet{Furlan08} to include the sheet-collapse models developed by \citet[1994,][]{Hartmann96}.  The sheet-collapse model uses the solution for the collapse of a self-gravitating, isothermal, infinite sheet initially in hydrostatic equilibrium; this model results in a relatively flat spectral energy distribution (SED) over the mid-infrared wavelength region. 


\begin{figure*}[htp]
\centering
\rotatebox{90}{\includegraphics*[width=0.40\textwidth]{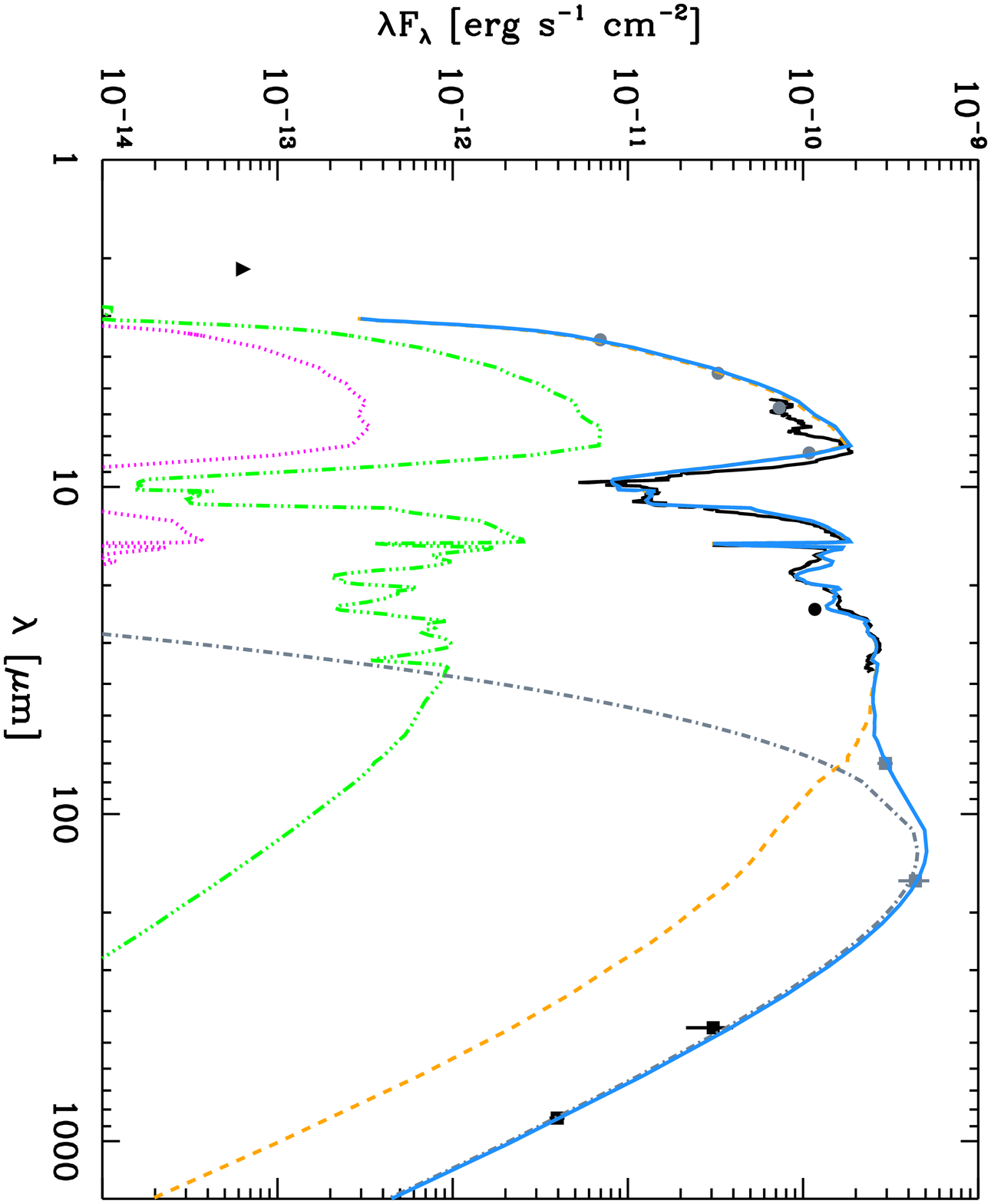}}\hspace{1.3mm}
\rotatebox{90}{\includegraphics*[width=0.40\textwidth]{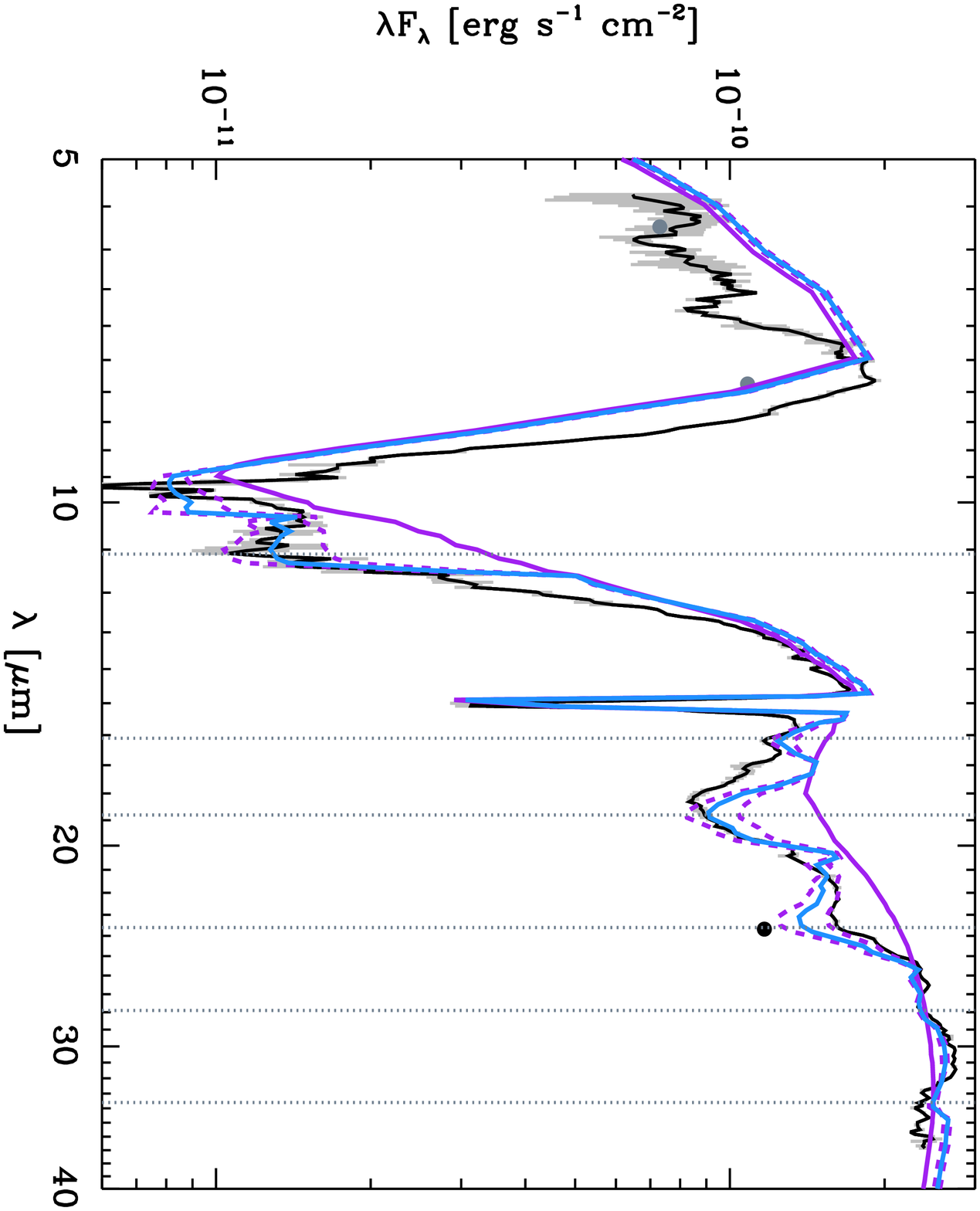}}
\caption{\emph{Left panel}:  Best-fit model compared to the observed SED of HOPS-68.  The composite SED includes the \emph{Spitzer}-IRS spectrum, PANIC \emph{K$_{s}$}-band flux (J.~J. Tobin, private communication), IRAC 3.6, 4.5, 5.8, and 8.0 $\micron$ and MIPS 24 $\micron$ fluxes (Megeath et al., in preparation), \emph{Herschel}-PACS 70 and 160 $\micron$ fluxes ($F_{\nu}$ = 6.9 and 23.3 Jy, respectively), and SCUBA 450 and 850 $\micron$ fluxes \citep{Johnstone99}.  The SCUBA estimates represent peak fluxes assuming a point-like source.  The model (\emph{blue line}) includes contributions from the central protostar (\emph{magenta line}), the disk (\emph{green line}), and the envelope (\emph{orange line}); the displayed disk and central protostar components are extinguished by the envelope.  In addition, a gray-body (\emph{dash-dotted line}) is also included to simulate external heating of the outer envelope.  \emph{Right panel}:  Close-up of the 5$-$40 $\micron$ region with IRS uncertainties (\emph{gray lines}) and forsterite wavelength positions indicated (\emph{dotted lines}).  Envelope models with increasing crystallinity (\emph{violet lines}; 0.0, 0.13, and 0.20, respectively) are shown for comparison against the best-fit model (\emph{blue line}; 0.17).  The mismatch near 8$-$9 $\micron$ between the adopted ``astronomical'' silicates and the observed amorphous silicate feature has been noted in the literature \citep{Henning10}.}
\label{fig:sed}
\end{figure*}


A central protostar and disk are responsible for heating the surrounding envelope.  The disk is modeled as a flat, irradiated, steady accretion disk that is optically thick at all wavelengths.  The most important input parameters include:  the total system luminosity ($L$ = $L_{\rm{star}}$+$L_{\rm{acc}}$), the fraction of the total luminosity that arises from the star ($\eta_{\rm{star}}$), the centrifugal radius ($R_{c}$), the inner ($R_{\rm{min}}$) and outer ($R_{\rm{max}}$) envelope radii, the inclination angle to the line-of-sight ($i$), the reference density at 1 AU ($\rho_{1}$), and the degree of asphericity \cite[$\eta$ = $R_{\rm{max}}/H$, where $H$ is the scale height of the original sheet; see][]{Hartmann96}.
  
We assume a similar envelope dust model as described in \citet{Osorio03} and \citet{Furlan08}.  The grain size distribution follows the standard power-law distribution $n$($a$)$da$ = $a^{-3.5}da$, with 0.005 $\micron$ $\leq$ $a$ $\leq$ 0.3 $\micron$.  The fractional abundance with respect to the mass of gas for the ice and dust species is similar to those proposed by \citet{Osorio03}:  $\zeta_{\rm{sil}}$ = 0.003,  $\zeta_{\rm{gra}}$ = 0.0025, $\zeta_{\rm{tro}}$ = 0.000768, $\zeta_{\rm{ice}}$ = 0.0012, and $\zeta_{\rm{co_{2}}}$ = 0.00039 for amorphous silicate, graphite, troilite, H$_{2}$O ice, and CO$_{2}$ ice grains, respectively. Details of the calculation and references for optical properties can be found in \citet{D'Alessio01}, \citet{Osorio03}, and \citet{Furlan08}.

In addition, we include a CDE grain shape distribution of forsterite in the Rayleigh-limit.  We adopt optical constants for (Mg, Fe)$_{2}$SiO$_{4}$ from \citet{Huffman73}, Mg$_{1.9}$Fe$_{0.1}$SiO$_{4}$ from \cite{Zeidler11}, Mg$_{2}$SiO$_{4}$ from \citet{Sogawa06}, and Mg$_{1.9}$Fe$_{0.1}$SiO$_{4}$ from \citet{Fabian01} for the 0.1$-$0.16, 0.24$-$2, 4$-$100, and 100$-$8000 $\micron$ wavelength regimes, respectively.  The optical constants from \citet{Fabian01} were modified to match the values from \citet{Sogawa06}; the real part $n$ was subtracted by a constant, while the imaginary part $k$ was multiplied by a scalar.  The MACs for forsterite were computed from the complex indices of refraction for each crystalline axis, and the three resultant values were then averaged together to account for randomly oriented grains \citep{Bohren83}.

The SED and best-fit model for \object{HOPS-68} are shown in the left panel of Figure \ref{fig:sed}.  Following the procedures of \citet{Furlan08}, the model parameters were adjusted by visual examination to yield a good fit over the entire SED with more emphasis given to the mid-infrared.  To account for the observed increase of flux beyond 70 $\micron$, we include a simple gray-body function \citep[\emph{dash-dotted line};][]{Gordon95} to simulate the effects of external heating on the outer envelope;  the dust temperature ($T_{d}$), emissivity index ($\beta$), and optical depth at 100 $\micron$ ($\tau_{100}$) are determined using a nonlinear least-squares fitting routine.  The best-fit parameters are tabulated in Table \ref{tbl:parameters}.

\begin{deluxetable}{lc}
\centering
\tablewidth{0.65\hsize}
\tablecolumns{2}

\tablecaption{Best-Fit Model Parameters \label{tbl:parameters}}

\tablehead{
\colhead{Parameter} &
\colhead{Value} \\
\cline{1-2}\vspace{-0.05 in} \\
\multicolumn{2}{c}{Sheet-Collapse}
}

\startdata
$L$ [$L_{\sun}$] \dotfill  & \phantom{{$.5$}}1.3 \\
$\eta_{\rm{star}}$ \dotfill & \phantom{{$.5$}}0.3 \\
$\eta$ \dotfill & \phantom{{$5$}} 2.0 \\ 
$R_{c}$ [AU] \dotfill & \phantom{{$.5$}}0.5 \\
$R_{\rm{min}}$ [AU] \dotfill & \phantom{$.55$}0.39 \\
$R_{\rm{max}}$ [AU] \dotfill & 7000\phantom{$555$} \\
$\rho_{1}$ [10$^{-14}$ g cm$^{-3}$] \dotfill & \phantom{$.5$}5.7 \\
$i$ [deg] \dotfill & 41\phantom{$5$} \\
$\theta$ [deg]\tablenotemark{a} \dotfill & 18\phantom{$5$} \\
$\zeta_{\rm{for}}$/($\zeta_{\rm{for}} + \zeta_{\rm{sil}}$) \dotfill & \phantom{$.55$}0.17 \\
\cutinhead{Gray-Body}
$T_d$ [K] \dotfill & \phantom{$.$}21.5 \\
$\tau_{100}$ \dotfill & \phantom{$.555$}0.012 \\
$\beta$ \dotfill & \phantom{$.55$}1.17 \\
$\chi_{\rm{red}}^{2}$ \dotfill & \phantom{$555$} 0.5
\tablenotetext{a}{Cavity semi-opening angle.}
\enddata

\end{deluxetable}

The SED modeling indicates a moderately luminous (1.3 $L_{\sun}$) protostar with a flattened ($\eta$ = 2.0), relatively high density envelope ($\rho_{1}$ = 5.7 $\times$ 10$^{-14}$ g cm$^{-3}$) and a small centrifugal radius ($R_{c}$ = 0.5 AU), and is viewed at an inclination of 41$\degr$.  \citep[Assuming a central mass of $M_{\ast}$ = 0.5 $M_{\sun}$, the reference density suggests a mass infall rate of \emph{\.{M}} = 7.6 $\times$ 10$^{-6}$ $M_{\sun}$ yr$^{-1}$;][]{Kenyon93}.  A small centrifugal radius may be the result of a very slowly rotating molecular cloud prior to collapse or may indicate a very short elapsed time since the onset of collapse.

The best-fit SED model has a forsterite mass fractional abundance of $\zeta_{\rm{for}}$ = 0.00063 relative to gas and adequately fits the depth of the 11.1, 16.1, 18.8, 23.6, 27.9, and 33.6 $\micron$ forsterite features (right panel of Figure \ref{fig:sed}).  The best-fit mass fractional abundances for crystalline and amorphous silicates ($\zeta_{\rm{for}}$/($\zeta_{\rm{for}}+\zeta_{\rm{sil}}$) = 0.00063/0.00363) imply a crystalline mass fraction of 0.17 ${}^{+3}_{-4}$.  This result is consistent with the cold absorbing screen method; the slightly larger mass fractions from the radiative transfer model may result from forsterite emission features generated in the warm ($\gtrsim$ 100 K) inner envelope, which must then be absorbed by the cooler outer envelope.

\section{Discussion \label{sec:discus}}

We have identified forsterite (Mg$_{2}$SiO$_{4}$) absorption in the cold, infalling envelope of a protostar, and estimate a crystalline mass fraction of 0.14 and 0.17 from the cold absorbing screen and radiative transfer methods, respectively.  These values are three to four times greater than the upper limit found in the interstellar medium \citep[$\lesssim$ 0.02-0.05;][]{Kemper05,Li07}.  The good agreement between the crystalline mass fractions, determined using two different methods with different optical constants, demonstrates the robustness of this result.  Even if we adopt the discrepant mass fraction of 0.06 estimated from the 18.8 and 23.6 $\micron$ forsterite features (Table \ref{tbl:massfrac}), the observed crystalline mass fraction still exceeds the upper limit for the interstellar medium.

Crystalline silicates are observed in emission toward the disks of young stars \citep[e.g.,][]{Sargent09}, where the temperature in the inner disk region is adequate for annealing; however, it is unlikely that the observed forsterite absorption arises in the disk.  While absorption features are produced in the atmospheres of rapidly accreting disks, such observations typically show features longward of 10 $\micron$ in emission \citep{Green06}.  Furthermore, since the contribution of the envelope is more than a factor-of-ten greater than the attenuated disk component (Figure \ref{fig:sed}), an absorption feature in the disk would require an unphysical depth to produce an observed depth of 20\% in the combined SED.

If the envelope is primarily material transported from the interstellar medium through gravitational collapse, what process is responsible for producing the observed crystalline grains?  For a slowly rotating, flattened envelope, much of the infalling material is deposited close to the central protostar, where it is then readily heated to temperatures sufficient for annealing.  But, because forsterite absorption is detected in the 20$-$30 $\micron$ regime, the absorbing grains must be located in a relatively low temperature ($\lesssim$ 100 K) region ($r$ $\gtrsim$ 15 AU).  It is highly unlikely that the amorphous silicates are annealed \emph{in situ} by irradiation, since raising the dust temperatures at distances $\gtrsim$ 15 AU above the glass transition temperature \citep[$T_{\rm{glass}}$ = 990~K;][]{Speck08} requires luminosities $\gtrsim$ 10$^{4}$ $L_{\sun}$.  Therefore, an alternative scenario for the production of crystalline silicates in the outer envelope is needed.

We propose that the amorphous silicates were annealed within the hot ($\gtrsim$ 1000 K) inner disk and/or envelope regions ($\lesssim$ 0.4 AU) and subsequently transported outward into the surrounding cold envelope material by entrainment in a protostellar outflow.  (An outflow associated with the protostar was detected by \citet{Williams03}.)  Alternatively, the amorphous silicates in the envelope may have been annealed \emph{in situ} by outflow-driven shocks; however, it is not clear that such shocks could heat grains to 1000 K without destroying them via sputtering or gain-grain collisions \cite[e.g.,][]{Draine93,Neufeld94}.

If these mechanisms result in short annealing times in which chemical equilibrium is not achieved, they would produce the observed dominance of forsterite over enstatite \citep{Gail04}.  While further analysis is needed to better understand the origin of the observed forsterite, its presence suggests that crystalline silicates detected in the outer disks of young stars may have been deposited during the protostellar infall phase.

\begin{acknowledgements}

The \emph{Spitzer Space Telescope} is operated by the Jet Propulsion Laboratory, California Institute of Technology, under NASA contract 1407.  This work was supported by NASA through contract number 1289605 issued by JPL/Caltech.  C.~A.~P. thanks Henrik Spoon and Adwin Boogert for insightful discussions. \\ \\

\end{acknowledgements}




\end{document}